\documentclass[12pt]{article}
\usepackage{pic02}
\usepackage{hyperref}
\usepackage{url}
\usepackage{graphicx}

\begin{document}

\title{\bf TOP AND HIGGS PHYSICS AT THE TEVATRON}
%\textit{PHYSICS IN COLLISION} PROCEEDINGS. UPPER-CASE, BOLD-FACE}
\author{Pierre Savard \\
{\em University of Toronto and TRIUMF}}
\maketitle

%
% photograph of author
%  This is where we will insert a photograph. To see what it would look like,
%  uncomment the following lines.
%
%\begin{figure}[h]
%\begin{center}
%
% include photograph for proceeding version
%
%\includegraphics[height=4.5cm]{einstein.eps}
%
% insert a fixed vertical spacing instead for the ArXiv preprint
%
\vspace{4.5cm}
%
%\end{center}
%\end{figure}

\baselineskip=14.5pt
\begin{abstract}

  We present a summary of our experimental understanding of
the top quark and discuss the significant improvements
expected in Run II at the Fermilab Tevatron Collider.  
We also discuss prospects for a Higgs boson discovery at
the Tevatron.

\end{abstract}
\newpage

\baselineskip=17pt

\section{Introduction}

  One of the great unsolved puzzles in particle physics
revolves around the generation of mass: how do particles 
become massive?  this question is answered 
in the context of the Standard Model by the Higgs mechanism,
which is evoked to break electroweak symmetry.  It predicts
the existence of a neutral scalar particle, the yet unobserved 
Higgs boson.

  In the Standard Model, the coupling strength between a fermion 
and the Higgs boson, the Yukawa coupling, determines the fermion
mass, which is not predicted by the theory.  For all fermions,
this coupling is small, except for one particle: the top quark.  
The top quark mass is close to the electroweak breaking scale
and its Yukawa coupling is within a few percent of unity.
Those intriguing characteristics have lead some to speculate that
the top quark itself is involved in electroweak symmetry breaking
or that its mass is a fundamental parameter of a more basic, 
underlying theory.  

  We currently know experimentally very little about both the 
Higgs boson and the top quark.  However, in the next few years, the CDF 
and D$\emptyset$ experiments at the Fermilab Tevatron Collider will pursue
a program of precision measurements of the top quark and will conduct 
an intensive search for the Higgs boson.   

In these proceedings, I will summarize our current experimental 
understanding of the top quark and discuss the prospects
for precision measurements in the coming years.  I will then
give an overview of the strategies that will be used to 
search for the Higgs boson.

\section{The Top Quark}

The top quark was discovered by the CDF and D$\emptyset$ collaborations 
in 1995\cite{discov}.  
This discovery was not unexpected since the structure of the SM requires
that there be an SU(2) partner to the b quark.  What did come as a surprise 
was the top quark's mass: 175 $GeV/c^2$, about 35 times more massive than 
the b quark, the heaviest known fermion at the time. Given its mass, many
properties of the top quark can be calculated within the Standard Model and are
given in Table~\ref{SMtop}.   

  An interesting consequence of the very short lifetime of the top quark is that it 
decays before it has time to hadronize. There are therefore no top hadrons and
the top quark can thus be studied as a ``free'' quark.

\begin{table}
\centering
\caption{ \it Top quark in th SM
}
\vskip 0.1 in
\begin{tabular}{|l|c|} \hline
          & simulation \\
\hline
\hline
 charge  & +2/3e   \\
  spin  & 1/2 \\
 width  & 1.4 $GeV/c^2$ \\
 mass  &  175 $GeV/c^2$ \\
 $BR (t \rightarrow Wb)$ &  $\sim$ 100\% \\
 lifetime  &  $\sim 5 \times 10^{-25} s$ \\
\hline
\end{tabular}
\label{SMtop}
\end{table}

\subsection{Run 1 measurements}

\begin{figure}[htb]
\includegraphics[width=13cm]{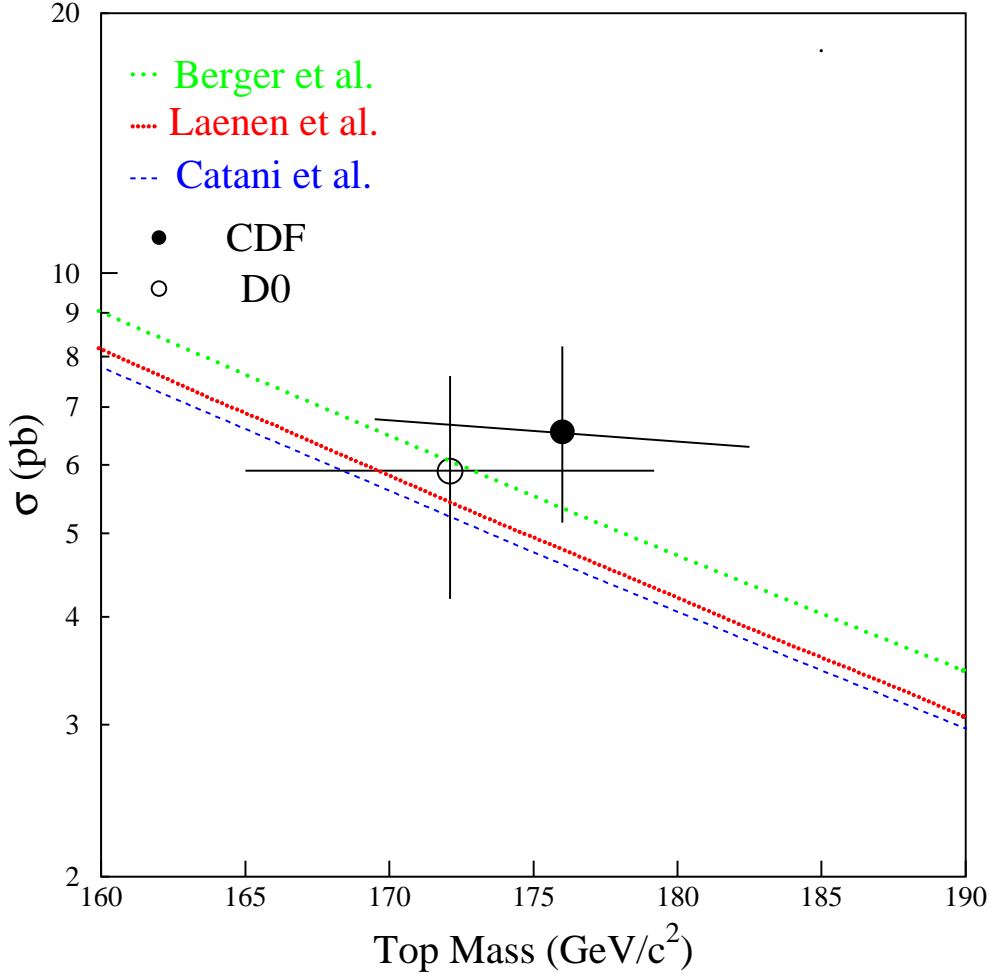}
 \caption{\it
      Example of a single figure with caption.
    \label{topxsmass} }
\end{figure}

 After the discovery of the top quark, the CDF and D$\emptyset$ collaborations
undertook studies aimed at characterizing its properties.  Because
of the statistically limited data sample available in Run 1, most measurements 
had large uncertainties, with the notable exception of the top mass (see 
Table~\ref{run1top} and Figure~\ref{topxsmass}).
Therefore, our current experimental understanding of the top quark is still
very limited and leaves room for potential surprises.

\begin{table}
\centering
\caption{ \it Some Run I Measurements}
\vskip 0.1 in
\begin{tabular}{|l|c|} \hline
          & simulation \\
\hline
\hline
 $BR  (t \rightarrow Zq) $ & $<$ .33 at 95\% CL  (CDF~\cite{fcnc}) \\
 $BR (t \rightarrow \gamma q) $ & $<$ .03 at 95\% CL  (CDF~\cite{fcnc}) \\
 $BR ( t \rightarrow H b) $  & $<$ .36 at 95\% CL\footnote{Higgs mass $<$ 160 $GeV/c^2$ and}  (D$\emptyset$~\cite{thb}) \\
 {\large ${BR  (t \rightarrow W b)} \over {BR  (t \rightarrow W q)}$ } & $>$ .56 at 95\% CL (CDF~\cite{chia})  \\
 $t\bar{t}$ spin correlation   & $\kappa$ $>$ -.25 at 68\% CL  (D$\emptyset$~\cite{spincor}) \\
\hline
\end{tabular}
\label{run1top}
\end{table}

\subsection{Run II Top Physics Program}

  With the start of Run II at the Tevatron, the era of precision top 
physics has begun.   During this period that started in 2000 and that 
should extend until the start of the Large Hadron Collider (LHC) at CERN, 
we expect to collect a clean data sample of over 1000 events where
at least one B hadron is identified.  Table~\ref{goals} lists some measurement goals
from the CDF and D$\emptyset$ experiments~\cite{d0tdr,cdftdr}.  In the following, I will
focus on two of the most important top physics measurements in Run II.

\begin{table}
\centering
\caption{ \it Some Top Physics Goals for 2 fb$^{-1}$
}
\vskip 0.1 in
\begin{tabular}{|l|c|} \hline
          &  uncertainty\\
\hline
\hline
 mass  & 2-3 $GeV/c^2$   \\
 $t\bar{t}$ cross section  &  9\% \\
 $V_{tb}$  & 13\% \\
 $BR  (t \rightarrow W b) $& 3\% \\
 $BR  (t \rightarrow Z q) $& $< 1 \times 10^{-2}$ \\
 $BR  (t \rightarrow \gamma q) $& $< 3 \times 10^{-3}$ \\
 $BR$  (W longitudinal) & 6\% \\
\hline
\end{tabular}
\label{goals}
\end{table}

\subsubsection{Single top production} 

The top quark discovery relied on the production of $t\bar{t}$ pairs via the strong interaction.
Top quarks can also be produced singly, via the electroweak interaction.  At the Tevatron,
two production processes dominate: the s-channel W* process, and the t-channel 
W-gluon fusion process\cite{zack}

  The single top production cross section is proportional to the CKM matrix element ${V_{tb}}^2$.
The measurement of the cross section can thus be interpreted as a measurement of $V_{tb}$,
with fewer assumptions than in branching ratio measurements that can be used to 
infer $V_{tb}$\cite{chia}.  The production cross section can also be sensitive to non-standard 
couplings of the  top quark, making it a very interesting observable to search 
for new physics\cite{tait}.  

  Understanding single top production will also be important for the Tevatron
Higgs search since the $WH$ channel, with $H \rightarrow b\bar{b}$,  has
an identical final state with kinematics very similar to the $W^{\star}$ channel.

  Finally, because the single top production proceeds through the V-A interaction, 
the resulting top quarks are polarized.  Since top quarks decay before they hadronize,  
we should be able to make top quark polarization measurements that have only 
been made with leptons.

\subsubsection{Top mass measurement}

 Precision electroweak measurements have allowed us to probe the
effect or radiative corrections on various electroweak observables.
Those radiative corrections can exhibit a quadratic dependence on the top 
quark mass, making this parameter and its experimental error important 
inputs to global electroweak fits.  In fact,  our current understanding of the 
self-consistency of the Standard Model and our knowledge of the Higgs mass 
are limited by the accuracy to which we currently know the top quark mass\cite{teubert}.
For this reason, reducing the uncertainty on the top quark mass will be
one of the top priorities of the Run II physics program.

  The goal for the experiments in Run II is to reduce the uncertainty to
2-3 $GeV/c^2$.  Attaining this level of precision will require a large-scale
effort to significantly improve our understanding of the jet energy scale and
of hard initial and final state QCD radiation.  Part of the needed improvements 
will  be made possible with the use of {\it in situ} calibration samples e.g. 
the $W \rightarrow jj$ resonance present in $t\bar{t}$ decay or
the $Z \rightarrow b\bar{b}$ resonance.

\section{Higgs Search at the Tevatron}

 In the next few years, the Tevatron experiments will have a unique opportunity 
to find or exclude the SM Higgs boson before the LHC experiments are 
commissioned.  

After a short review of the production and decay properties of the Higgs
boson, I will give a summary of the strategies and techniques that will be 
employed in this ambitious search. A detailed study can be found in 
the Report of the Tevatron Higgs Working Group\cite{higgs}.

\subsection{Higgs Production and Decay}

\begin{figure}[htb]

\includegraphics[width=11cm,angle=270]{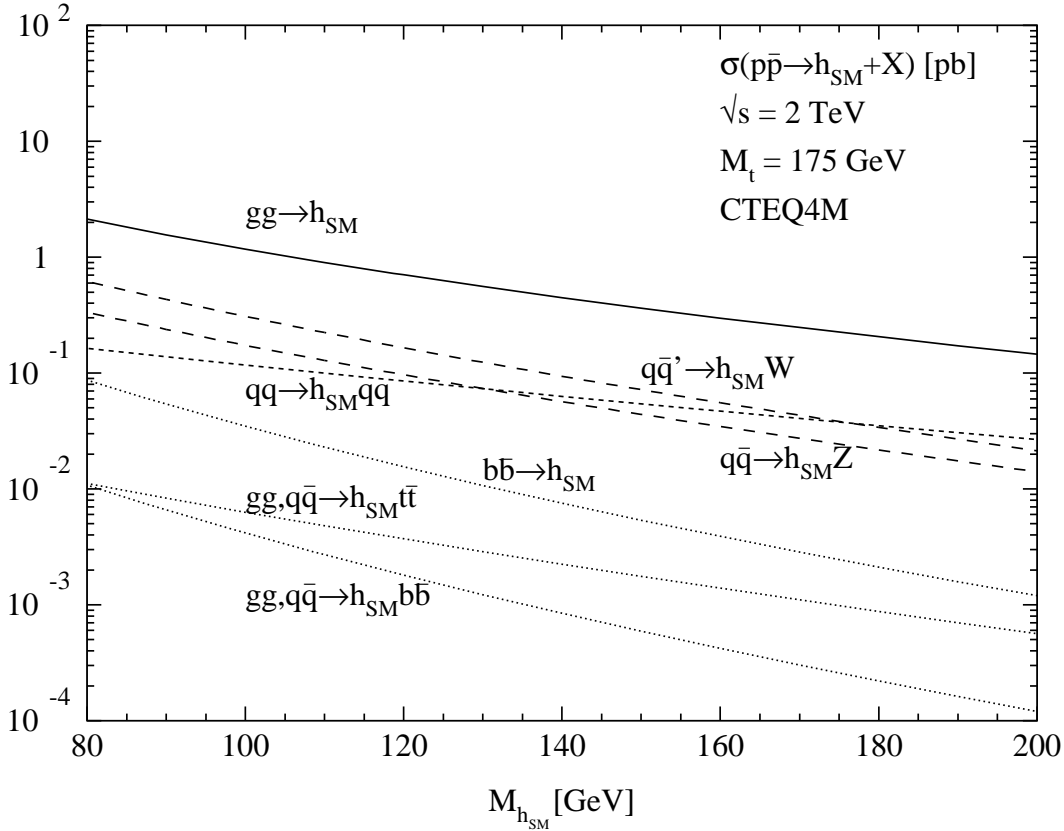}
 \caption{\it
            Higgs production cross section versus Higgs mass for various
   production processes.
  \label{higgsxs} }
\end{figure}

  Figure~\ref{higgsxs} shows the Higgs production
cross section at the Tevatron as a function of the Higgs boson
mass.  Over the mass range shown, the gluon-gluon fusion process
dominates. The cross section for WH and ZH production is lower, but
these modes allow for smaller backgrounds and a straightforward trigger 
signature through the leptonic decay of the weak bosons.  The
$Ht\bar{t}$ mode has a spectacular decay signature with two W bosons
and 4 b quarks but suffers from a very small cross section.

\begin{figure}[htb]
\includegraphics[width=14cm]{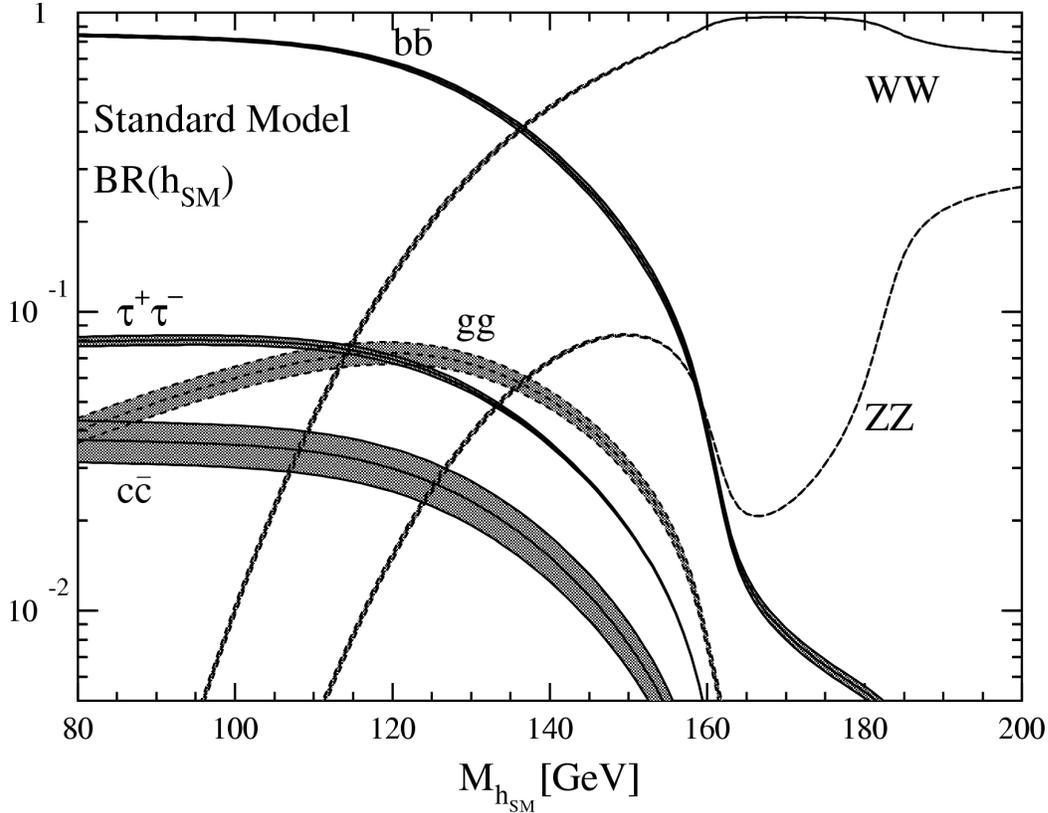}
 \caption{\it   Higgs branching ratios versus Higgs mass.
    \label{higgsbr} }
\end{figure}

  Figure~\ref{higgsbr} shows the Higgs branching ratio as a
function of Higgs mass.  Two mass ranges can be identified: the low-mass
range where the dominant Higgs decay mode is to a pair of b quarks and the
high-mass range where the Higgs decays to a pair of W bosons.  Different
experimental strategies have been developed to deal with these two mass
regimes.

\subsubsection{Experimental tools}

 As mentioned above, the  low-mass regime is dominated by Higgs
decays to $b\bar{b}$.  In order to maximize the search sensitivity,
we will need to optimize B tagging efficiency and improve dijet 
mass resolution.

To observe a $H \rightarrow b\bar{b}$ resonance that is as
narrow as possible in the background-dominated dijet mass spectrum, 
we need to make improvements in the jet energy resolution.  To this
effect, CDF developed in Run I a technique that makes use of tracking,
preshower, and shower maximum detectors to improve its energy
resolution.  It is hoped that both detectors will be able to
achieve a $H \rightarrow b\bar{b}$ mass resolution of about 10\%.

  The two experiments have much improved tracking systems for
Run II.  The D$\emptyset$ detector now has a large silicon vertex detector
surrounded by a fiber tracker.  The two devices are immersed in a 2 Tesla 
magnetic field. CDF replaced its drift chamber and its silicon vertex
detector which now has an acceptance that extends to $| \eta | < 2$.  These
improvements will translate into significantly higher b-tagging 
efficiencies relative to what was obtained in Run I.  The results that
will be presented later assumed a tight b-tag efficiency of 60\% and a
loose b-tag efficiency of 75\%, while keeping the fake rate under 1\%. 

 The Tevatron Higgs Working Group studied potential improvements that
could be made by using neural networks, which have been extensively 
used by the LEP experiments in their Higgs searches.  The outcome
of the study was very promising and a neural net analysis was
used to obtain the results presented in Figure~\ref{final}.

  Before the Higgs search attains the sensitivity required for
an observation, we will be searching for two Standard Model signals 
that have not yet been observed: single top in the $W^{\star}$ 
channel and $WZ$, with $Z \rightarrow b \bar{b}$.  Both these 
channels have the same final state with similar kinematics 
as the $WH$ channel but have significantly higher $\sigma \times BR$.  
This will provide opportunities to exercise analysis 
tools and estimate background rejection techniques well before 
a Higgs signal becomes visible.

\subsection{Results from Tevatron Higgs Working Group}

\subsubsection{Low Mass Range}

  Results from the $WH$ and $ZH$ channels with $H \rightarrow b\bar{b}$ and the leptonic decay
of the weak bosons are presented in Table~\ref{low}.  One jet from the Higgs decay is
required to be tagged by a ``tight'' algorithm and the other jet by a ``loose'' 
algorithm whose efficiencies were mentioned earlier.
The main backgrounds come from $t\bar{t}$, $Wb\bar{b}$, single top, and WZ.    

\subsubsection{High Mass Range}

  The high mass region is dominated by the Higgs decay to
$W W^{\star}$ and $WW$.  In this case, one can afford to make use
of the gluon fusion production mechanism that has a higher cross section.
However, the results below also use the associated Higgs
production mechanism since it can provide striking trilepton signatures.
The main backgrounds come from $WW$, $WZ$, $ZZ$, and $t\bar{t}$. The results
of the study are presented in Table~\ref{high}.

 \begin{table}
   \begin{center}
   \caption{Low-mass Higgs search
                 sensitivities per detector in 1 fb$^{-1}$ }
\vskip 0.1 in

    \label{low}
%%   \begin{tabular}{|c|c|ccccc|} \cline{3-7}
%%    \multicolumn{2}{l}{ }           & \multicolumn{5}{|c|}{Higgs mass (GeV/$c^2$)}  \\  
%\setlength{\tabcolsep}{1pc}
     \begin{tabular}{llccccc}  \hline
                      &              & \multicolumn{5}{c}{Higgs Mass (GeV/$c^2$)} \\ \hline \hline
\noalign{\vskip3pt}
\multicolumn{1}{c}{Channel} &\multicolumn{1}{c}{Rate}& 
                                        90   &  100  &  110  &  120  &  130 \\ 
%--------------------------------------------------------------------------------------
                      & $S$          &  8.7  &  9.0  &  4.8  &  4.4  &  3.7  \\  
 $\ell\nu b\bar{b}$     & $B$          &  28   &  39   &  19   &  26   &  46   \\ 
                      & $S/\sqrt{B}$ &  1.6  &  1.4  &  1.1  &  0.9  &  0.5  \\  \hline  \hline
%--------------------------------------------------------------------------------------
                      & $S$          &  12   &   8   &  6.3  &  4.7  &  3.9  \\   
 $\nu\nu b\bar{b}$         & $B$          & 123   &  70   &  55   &  45   &  47   \\  
                      & $S/\sqrt{B}$ &  1.1  &  1.0  &  0.8  &  0.7  &  0.6  \\  \hline \hline
%--------------------------------------------------------------------------------------
                      & $S$          &  1.2  &  0.9  &  0.8  &  0.8  &  0.6  \\    
 $\ell \ell b\bar{b}$      & $B$          &  2.9  &  1.9  &  2.3  &  2.8  &  1.9  \\   
                      & $S/\sqrt{B}$ &  0.7  &  0.7  &  0.5  &  0.5  &  0.4  \\  \hline \hline
%--------------------------------------------------------------------------------------
    \end{tabular} 
      \end{center} 
  \end{table}

\begin{table}
   \begin{center}
   \caption{High-mass Higgs search
                 sensitivities per detector in 1 fb$^{-1}$ }
   \label{high}

\vskip 0.1 in
 
%  \begin{tabular}{|c|c|ccccccc|} \cline{3-9}
%     \multicolumn{2}{l}{ }           & \multicolumn{7}{|c|}{Higgs mass (GeV/$c^2$)}  \\
\setlength{\tabcolsep}{9pt}
  \begin{tabular}{llccccccc}   \hline
                      &              & \multicolumn{7}{c}{Higgs Mass (GeV/$c^2$)}  \\ \cline{3-9}
\noalign{\vskip3pt}
\multicolumn{1}{c}{Channel} & \multicolumn{1}{c}{Rate}  &
                                         120  &  130  &  140  &  150  &  160  &  170  &  180  \\  
%------------------------------------------------------------------------------------------------------
%                      & $S$          & 0.011 & 0.025 & 0.039 & 0.050 & 0.057 & 0.033 & 0.033 \\      
% $\ell \ell \ell$            & $B$          & 0.025 & 0.025 & 0.025 & 0.025 & 0.025 & 0.025 & 0.025 \\    
%                      & $S/\sqrt{B}$ & 0.07  &  0.16 &  0.25 &  0.32 &  0.36 &  0.21 &  0.21 \\  
%------------------------------------------------------------------------------------------------------
                      & $S$          &   -   &   -   &  2.6  &  2.8  &  1.5  &  1.1  &  1.0  \\      
 $\ell \ell \nu \nu$            & $B$          &   -   &   -   &  44   &  30   &  4.4  &  2.4  &  3.8  \\    
                      & $S/\sqrt{B}$ &   -   &   -   &  0.39 &  0.51 &  0.71 &  0.71 &  0.51 \\  \hline
%------------------------------------------------------------------------------------------------------
                      & $S$          &  0.08 &  0.15 &  0.29 &  0.36 &  0.41 &  0.38 &  0.26 \\      
 $jj \ell \ell$           & $B$          &  0.58 &  0.58 &  0.58 &  0.58 &  0.58 &  0.58 &  0.58 \\    
                      & $S/\sqrt{B}$ &  0.11 &  0.20 &  0.38 &  0.47 &  0.54 &  0.50 &  0.34 \\  \hline
%------------------------------------------------------------------------------------------------------
   \end{tabular}
   \end{center}
 \end{table}

\begin{figure}[htb]
\includegraphics[width=15.0cm]{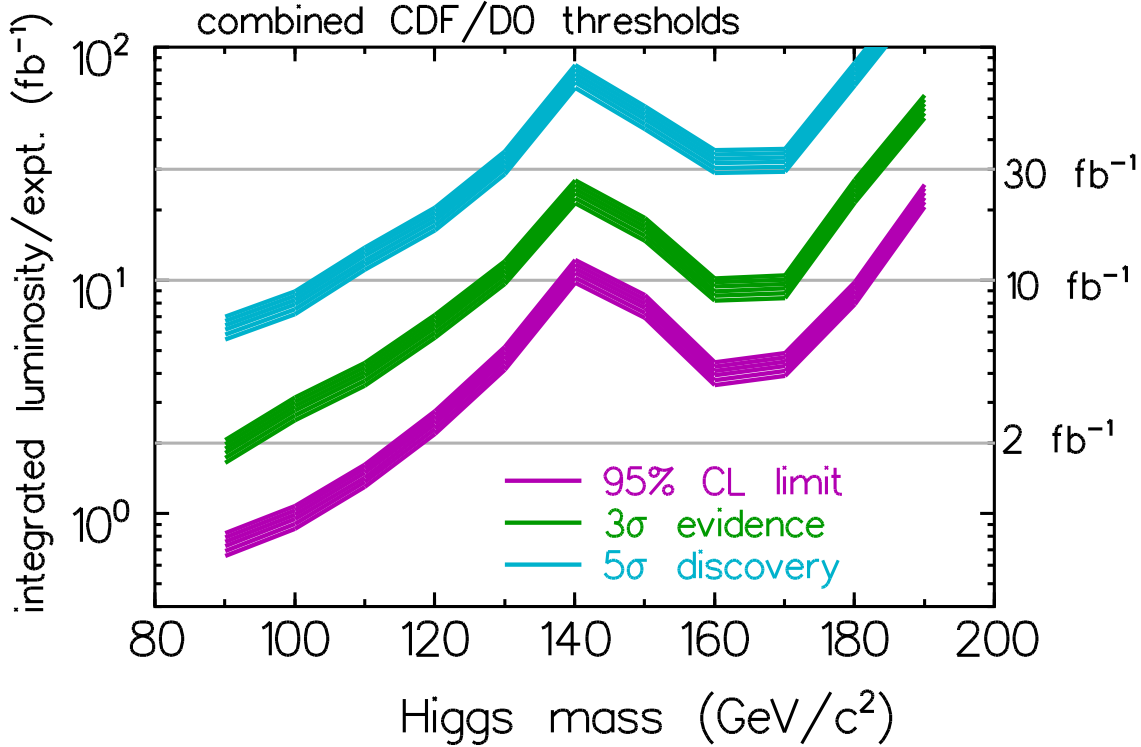}
 \caption{\it
      Integrated luminosity required to observe or exclude a
 Higgs boson versus Higgs mass.
    \label{final} }
\end{figure}

\subsubsection{Combined Results}

The integrated luminosity required to observe or exclude
a Higgs Boson as a function of Higgs mass is shown in Figure~\ref{final}.  
The study finds that a SM Higgs with a mass up to 135 $GeV/c^2$
can be excluded at 95\% CL with 6 $fb^{-1}$ and up to 
180 $GeV/c^2$ with 10 $fb^{-1}$. With 15 $fb^{-1}$, one could
obtain 3 $\sigma$ ``evidence''  up to 135 $GeV/c^2$.

 The study assumed a $b\bar{b}$ dijet mass resolution of 10\%,
b-tagging efficiencies of 75\% on one of the B hadrons (tight 
algorithm), and 60\% on the other (loose algorithm).  Acceptance 
and efficiencies for signal and backgrounds were estimated using a 
Monte Carlo simulation that parametrized the response of 
a generic Run II detector.  The size of the bands on Figure~\ref{final} reflect a
30\% uncertainty that includes b-tagging, mass resolution, 
background rate and other effects.

\section{Conclusion}

  With the beginning of Run II at the Tevatron, the era
of precision top physics has begun and the hunt for the
Higgs boson has resumed.  This exciting period
will be followed with the commissioning of the LHC, a 
true ``top quark factory''.  The measurements made in the
next decade should allow us to determine whether the generation
of mass for elementary particles is indeed described 
by the Higgs mechanism of the Standard Model and whether
the top quark plays a special role in the breaking of
electroweak symmetry.

%\section{Acknowledgements}

%
\end{document}